\documentclass[aps,superscriptaddress,groupedaddress]{revtex4}  
\usepackage{graphicx}  
\usepackage{dcolumn}   
\usepackage{bm}        
\usepackage{amssymb}   
\usepackage{siunitx}
\usepackage{blindtext}
\usepackage{inputenc}
\usepackage{amsmath}
\usepackage{amssymb}
\usepackage{graphicx}
\usepackage{natbib}
\usepackage{lipsum}
\usepackage{mathrsfs}
\usepackage{color}
\usepackage[abs]{overpic}

\begin{document}
\title{Quantum magnetomechanics: towards the ultra-strong coupling regime}

\author{E. Romero-S\'anchez}
\affiliation{ARC Centre for Engineered Quantum Systems, School of Mathematics and Physics, The University of Queensland, Brisbane, Queensland 4072, Australia}
\author{W. P. Bowen}%
\affiliation{ARC Centre for Engineered Quantum Systems, School of Mathematics and Physics, The University of Queensland, Brisbane, Queensland 4072, Australia}
\author{M. R. Vanner}
\affiliation{ARC Centre for Engineered Quantum Systems, School of Mathematics and Physics, The University of Queensland, Brisbane, Queensland 4072, Australia}
\affiliation{Clarendon Laboratory, Departament of Physics, University of Oxford, OX1 3PU, United Kingdom}
\author{K. Xia}%
\affiliation{ARC Centre for Engineered Quantum Systems, Department of Physics and Astronomy, Macquarie University, NSW 2109, Australia}
\affiliation{National Laboratory of Solid State Microstructures,
College of Engineering and Applied Sciences, Nanjing University, Nanjing 210093, China}

\author{J. Twamley}%
\affiliation{ARC Centre for Engineered Quantum Systems, Department of Physics and Astronomy, Macquarie University, NSW 2109, Australia}

\begin{abstract}
In this paper we investigate a hybrid quantum system comprising a mechanical oscillator coupled via magnetic induced electromotive force to an $LC$ resonator. We derive the Lagrangian and Hamiltonian for this system and find that the interaction can be described by a charge-momentum coupling with a strength that has a strong geometry dependence. We focus our study on a mechanical resonator with a thin-film magnetic coating which interacts with a nano-fabricated planar coil. We determine that the coupling rate between these two systems can enter the strong and ultra-strong coupling regimes with experimentally feasible parameters. This magnetomechanical configuration allows for a range of applications including electro-mechanical state transfer and weak-force sensing.
\end{abstract}

\maketitle

\section{Introduction}

Coupling between electromagnetic and mechanical degrees of freedom is central to a number of quantum science experiments and enables the development of many quantum technologies. Mechanical oscillators can act as coherent interfaces between different electromagnetic fields \cite{andrews2014bidirectional,
bagci2014optical} and are a promising tool for the development of future quantum technologies oriented to communications, memories and metrology. Additionally, due to their relatively large mass, mechanical systems offer a promising route to perform fundamental tests of quantum physics \cite{
caves1980measurement,
marshall2003towards,
pikovski2012probing}. A multitude of approaches in both opto- and electro-mechanics have been suggested and experimentally studied such as suspended mirrors forming an optical cavity with variable cavity length formed by 
microtoroids carrying whispering gallery modes \cite{carmon2005temporal}, $LC$ resonators with a mobile drum mode capacitor  \cite{palomaki2013entangling}, the motion of superfluid \cite{de2014superfluid,
harris2016laser} and nano-phononic crystals \cite{eichenfield2009optomechanical}. 

The basic coupling in optomechanics and electromechanics is fundamentally similar but physically different. In both cases a mechanical displacement produces a shift in the resonance frequency of an electromagnetic resonator. In optomechanics optical resonators are formed by mobile elements that change the length of the cavity. In electromechanics, capacitors used in $LC$ circuits are commonly formed by one mobile plate, so the resonance frequency is position dependent. Since the optomechanical coupling rate is related to the momentum transfer between the photon and a mechanical oscillator \cite{mcauslan2016microphotonic}, it is usually small, such that reaching beyond the strong coupling regime is complicated. In recent literature, the term optomechanics is used to refer to both opto- and electro-mechanical systems \cite{bowen2015quantum}, we follow this convention through this paper.

At the quantum level, many experimental control protocols require quantum-coherent exchange of excitations between the light and mechanical systems \cite{O’Connell2010}, which is possible when the optomechanical interaction is faster than the dissipation of the light and mechanics, known as strong coupling condition. Significant progress has been made in a variety of architectures that enables this strong coupling to be observed \cite{groblacher2009observation}. Strongly coupled systems have been used for instance, to cool down the state of motion of mechanical oscillators to their ground state \cite{teufel2011sideband} and the preparation of entangled states of motion of a macroscopic mechanical oscillator \cite{palomaki2013entangling}. The magnitude of the coupling rate defines two other main regimes that remain unexplored in optomechanics. The first one, referred to as ultra-strong coupling regime is accessible when the coupling rate is considerable fraction of the resonance frequency \cite{hu2015quantum,anappara2009signatures,rossatto2017spectral}. In optomechanical systems, the ultra-strong coupling regime has been proposed to exhibit novel physics at the quantum level \cite{holz2015suppression,rossatto2016entangling}.

Approaches that explore mechanical oscillators coupled to electric circuits through magnetic interactions have been referred to as \textit{magnetomechanics} and has been little explored compared to electromechanics \cite{ares2016resonant}. Quantum magnetomechanics explores different techniques to prepare and control quantum states of motion of a mechanical oscillator using magnetic interactions. Several approaches to quantum magnetomechanics have been proposed and include magnetically levitated mechanical oscillators with the aim of reduce decoherence \cite{romero2012quantum,cirio2012quantum}, and coupling the motion of a mechanical oscillator to a superconducting circuit \cite{via2015stromg}. The applications of quantum magnetomechanics can be expanded to systems with intrinsic magnetic properties such as electric circuits, superconducting qubits \cite{xia_opto-magneto-mechanical_2014} or spin qubits \cite{rabl2009strong}.

\begin{figure}[t]
\includegraphics[scale=0.125]{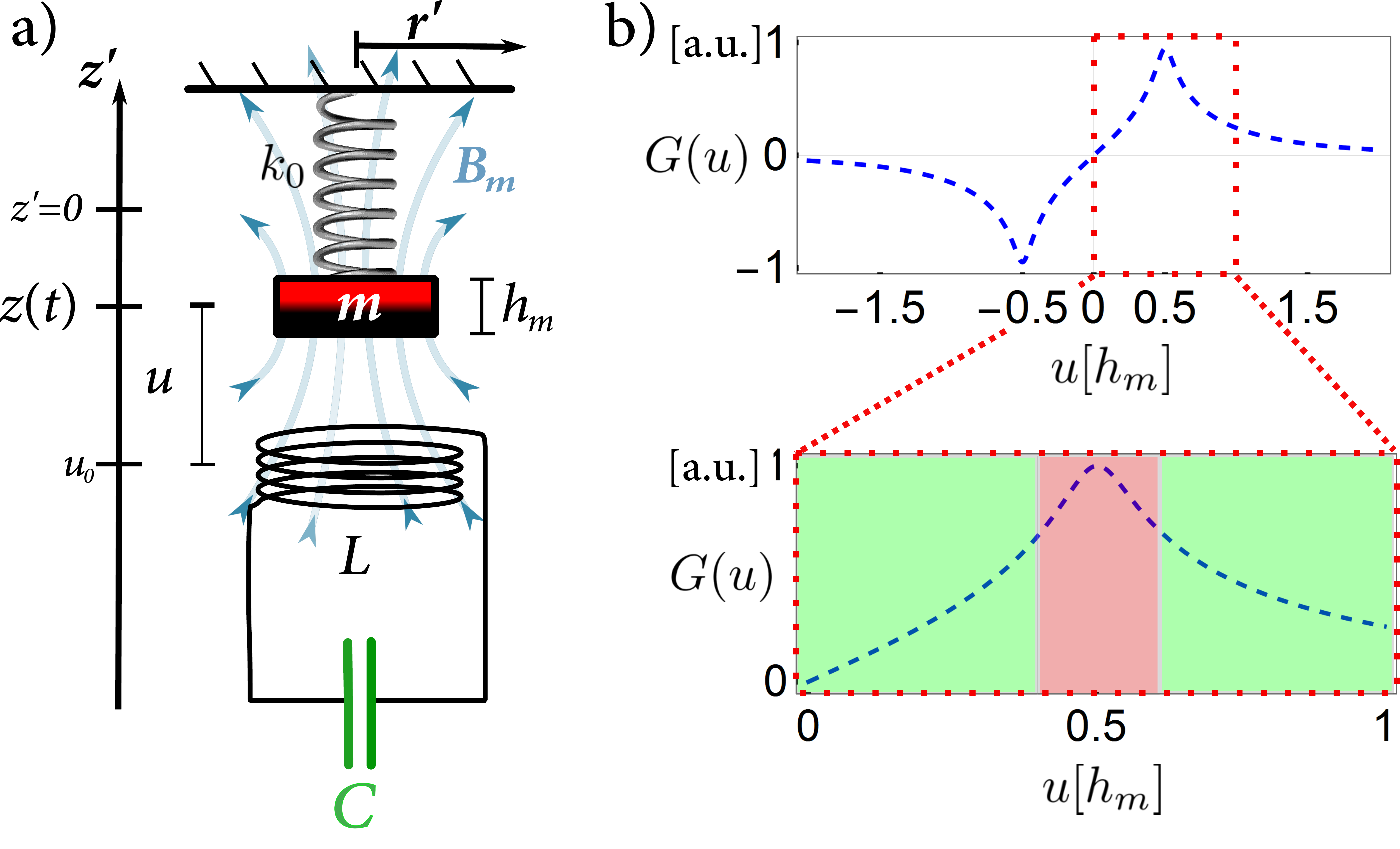}
\caption{a) Simplified scheme of the magnetomechanical system described in cylindrical coordinates $z'$ and $r'$. A cylindrical magnet of mass $m$ and thickness $h_{\textnormal{m}}$ attached to a spring of stiffness $k_0$ forms a mechanical oscillator. The mechanical oscillator is inductively coupled to an $LC$ resonator. The magnet produces a magnetic field ${\bf B}_{\textnormal{m}}$ that induces a flux in the inductor $L$ placed at $z'=u_0$, which is connected to a capacitor $C$. The equilibrium position of the center of mass of the magnet is $z'=u_0$ and it displaces $z(t)$ around it. b) Coupling rate $G(u)$ [a.u.] normalized to the maximum, which is proportional to the Faraday flux force on the magnet, as a function of the separation between the magnet and the coil $u$. The region where the linear (green) interaction is presented on both sides and where the dominant interaction is quadratic (red) is in the center. }\label{fig:LCScheme}
\end{figure}

In this paper we re-examine an electromechanical scheme, dating back as far as 1980 \cite{caves1980measurement} that utilizes inductive coupling, placing it on a solid theoretical formulation by deriving the Lagrangian and the associated Hamiltonian. We further explore experimental regimes that may be achieved using modern fabrication techniques. The magnetomechanical system that we study is composed of a mechanical oscillator coupled magnetically to an $LC$ resonator as shown in Fig. \ref{fig:LCScheme}a. In contrast to optomechanics, where the mediating force is due to the radiation pressure, in our magnetomechanical system the mediating force is the Lorentz force. Using micro/nano fabricated designs which are experimentally achievable we find that strong and ultra-strong coupling are attainable. 

An outline of the paper is as follows: in section \ref{Semiclassical} we determine the Lagrangian of the coupled mechanical-$LC$ system and then derive the quantized Hamiltonian. The quantum Hamiltonian exhibits either an adjustable linear or quadratic coupling (section \ref{QuantumDynamics}). In the linear coupling regime \ref{LinearRegime} we find that the $LC$ circuit couples to the mechanical momentum identical to the velocity sensor studied in \cite{caves1980measurement}. We explore the linear system in the strong coupling regime in section \ref{SC}, and the generalized dynamics beyond the strong coupling in section \ref{USC}. We find the energy spectrum and the eigenstates of the magnetomechanical system and estimate the occupation for its ground state, finding signs of entanglement between the mechanics and the electronics \ref{GS}.  We study the system in the presence of decoherence and analyse some of the spectral properties of the mechanical system in the presence of the inductive coupling in section \ref{LangevinEquations}. Properties of the proposed experimental device are discussed in section \ref{ExperimentalConsiderations} where we consider an implementation with high $Q$ mechanical oscillators and a superconducting $LC$ resonator. The linear dependence of the Hamiltonian on the mechanical momentum also suggest that the Hamiltonian is not invariant under the reversal of time i.e. $t\rightarrow -t$, and thus breaks time reversal symmetry and we discuss this in Appendix A.

\section{Semi-Classical Picture}\label{Semiclassical}

We define a cylindrical coordinate system $(r',z')$, where we consider a small cylindrical magnet with magnetization $M$ and vector ${\bf M}=M {\bf e}_{z'}$. The magnet generates a magnetic field ${\bf B}_{\textnormal{m}}(z',r')=B_z{\bf e}_{z'}+B_r{\bf e}_{r'}$ \cite{ravaud2010cylindrical}. In Fig. \ref{fig:LCScheme}a we represent a scheme for the magnet's position of the center of mass $z(t)$ its placed at its equilibrium position $z'=0$. The magnet, with effective mass $m$, is attached to a spring with spring constant $k_0$, forming a mechanical oscillator which resonates at the frequency $\omega_{\textnormal{m}}=\sqrt{k_0/m}$. The equation of motion for the center of mass of the magnet is given by $m \left[\ddot{z}(t) + \Gamma_{\textnormal{m}} \dot{z}(t) + \omega_{\textnormal{m}}^2 z(t)\right]=F_{\textnormal{ext}} (t)$, where $\Gamma_{\textnormal{m}}$ represents the mechanical damping rate and $F_{\textnormal{ext}} (z,t)$ is an arbitrary external driving force.

Here we use a planar electric coil with inductance $L$ placed vertically below the magnet at $z'=u_0$ and connected to a two plate capacitor $C$. This combination of capacitor and inductor creates a cavity with resonance frequency centred at $\omega_{\textnormal{e}}=1/\sqrt{LC}$. The equation of motion for an $LC$ resonator that is driven with an arbitrary external voltage $V_{\textnormal{ext}} (t)$ is given by $L\left[ \ddot{q}(t) + \Gamma_{\textnormal{e}} \dot{q}(t) + \omega_{\textnormal{e}}^2 q(t)\right]=V_{\textnormal{ext}} (t)$ where $\Gamma_e=R/L$ is the dissipation rate and $R$ is the resistance of the entire circuit.

The planar inductor follows a geometrical path in three dimensions, which we define via a vector path $\bf{S}$ whose transversal area element $d{\bf a}=d \text{a} \ {\bf e}_{z'}$ is normal to the plane where the inductor lays. The magnetic flux crossing the area enclosed by the inductor is $\Phi_B= \int {\bf B}_{\textnormal{m}}(u)\cdot d {\bf a}$, where $u=z(t)-u_0$ represents the relative vertical separation between the magnet and the coil. Here we treat the Lorentz force $F_L(t)$ as the dominant force acting on the magnet so $F_{\textnormal{ext}} (t)=F_L(t)$. and the electromotive force (EMF) $\mathcal{E}(t)$ as the main source of voltage $V_{\textnormal{ext}} (t)=\mathcal{E}(t)$. In section \ref{LangevinEquations} we discuss the case where the system is thermally driven.

As the magnet displaces along $z'$, it fluctuates around an equilibrium position $z'=u_0$. The mechanical vertical motion creates an AC magnetic field which couples the $LC$ circuit-mechanical system, via mutual inductance. The change in position induces then a change in flux generating an EMF in the electric circuit $\mathcal{E}(t)=-\frac{d \Phi_{B}}{d  t}$. The displacement is restricted to the $z'$ axis and it is parallel to the area component ${\bf a}$, using the expression for the time derivative of flux \cite{benedetto2016some}, the EMF can be re-expressed in terms of the magnetic field  
\begin{equation}\label{eq:EMF}
\mathcal{E}(t)=- \dot{z}(t)\oint_{\text{coil}} {\bf e}_{z'} \cdot \left( {\bf B}_{\textnormal{m}}(u) \times d{\bf S}\right).
\end{equation}

The induced $\mathcal{E}(t)$ produces an small current in the $LC$ circuit, and the inductor carrying the current generates a magnetic field which interacts with the magnetic field of the permanent magnet exerting a Lorentz force ${\bf {F}}_{L}(t)$ between the mechanical oscillator and the $LC$ circuit
\begin{equation}\label{eq:LorentzForce}
{\bf {F}}_{L}(t)= -\dot{q}(t) \oint_{\text{coil}} {\bf{B}}_{\textnormal{m}}(u) \times {d\bf{S}}.
\end{equation}
The Lorentz force ${\bf {F}}_{L}(t)=F_z(t){\bf e}_{z'}+F_r(t) {\bf e}_{r'}$ has a radial component $F_r(t) {\bf e}_{r'}$ which points radially inwards, therefore around a closed loop $\langle F_r(t)\rangle\approx 0$. The simplified expression for the Lorentz force is the contribution of the vertical component $ F_z(t)=-\dot{q}(t)\left( \oint_{\text{coil}} {\bf{B}}_{\textnormal{m}} (u) \times {d\bf{S}}\right)\cdot{\bf e}_{z'}$. The effect of this force acting on the mechanical oscillator produces a modification of the stiffness of the mechanical spring constant and we denote it as the Lorentz spring constant $k_L$.  By altering the current in the inductor this spring constant can be modified allowing one to electrically tune the mechanical resonance frequency. From Eq. \eqref{eq:EMF} and Eq. \eqref{eq:LorentzForce} we define 
\begin{equation}\label{eq:coupling}
G(u)= \oint_{\text{coil}}\left[{\bf B}_{m}(u) \times d{\bf S}\right]\cdot {\bf e}_{z'},
\end{equation}
as the magnetomechanical coupling term $G\propto M $, which couples the mechanical and electrical interactions through a magnetic interaction.  

The magnetomechanical coupling rate $G(u)$ is a function of the relative separation between the magnet and the coil $u=z(t)-u_0$. In Fig. \ref{fig:LCScheme}b we plot the coupling rate $G(u)$ as a function of $u$, which can be freely controlled in an experiment. We set the initial equilibrium separation $u_0$, with $u_0/h_{\textnormal{m}}\in [0,1]$ where $h_{\textnormal{m}}$ is the thickness of the magnet (Fig. \ref{fig:LCScheme}a). The small displacement of the magnet around $u_0$, allows us to expand $G(u)\rightarrow G(u_0+z(t))$ as a function of $z$, the canonical coordinate of the center of mass mechanical motion. For small displacements around $u_0$, we define the linear coupling rate $G_0=G(u_0)$ and expand $ G(z)\approx G_0+G_j z^{j}$ where the generalized expression $G_j=\partial^j G(z)/\partial z^{j}|_{z'=u_0}$ and $j=1,2$. The choice of $u_0$ will define two different regions that correspond to different dominant non-linear terms of $G(z)$. The first region is shown in color green in Fig. \ref{fig:LCScheme}b (bottom), the dominant interaction in this region is defined by the first order term ($j=1$). The second region is illustrated as a red coloured area in Fig. \ref{fig:LCScheme}b and the dominant non-linear term is the second order one ($j=2$). The following analysis is equivalent for $j=1,2$. For the sake of simplicity, we will focus on the interaction up to first order ($j=1)$.

Now we consider that the two oscillators are driven externally, with the force Eq. \eqref{eq:LorentzForce} for the mechanicas and the voltage Eq. \eqref{eq:EMF} for the electronics. The dynamics of the coupled system is then described by the set of coupled equations of motion 
\begin{equation}\label{eq:coupledequations}
\begin{array}{c}
m \left[\ddot{z}(t) + \Gamma_{\textnormal{m}} \dot{z}(t) + \omega_{\textnormal{m}}^2 z(t)\right]
=
- \dot{q}(t) G(z),
\\
\\
\\
L\left[ \ddot{q}(t) + \Gamma_{\textnormal{e}} \dot{q}(t) + \omega_{\textnormal{e}}^2 q(t)\right]
=
\dot{z}(t)G(z),
\end{array}
\end{equation}
clearly, the magnetomechanical system is coupled through $G(z)$. Keeping in mind that our goal is to obtain a quantum description of the system, we require the calculation of the lossless magnetomechanical Lagrangian $\mathscr{L}$ from which we can derive the equations of motion Eq. \eqref{eq:coupledequations}. We find that this lossless Lagrangian is
\begin{equation}\label{eq:Lagrangian}
\begin{array}{lc}
\mathscr{L}(z,q,\dot{z},\dot{q})
=&
\displaystyle\left(\frac{m}{2}\dot{z}^2-\frac{m}{2} \omega_{\textnormal{m}}^2 z^2\right)+\left(\frac{L}{2}\dot{q}^2-\frac{L}{2}\omega_{\textnormal{e}}^2 q^2\right)
\\
\\
& +
\displaystyle G(z)  z\dot{q}+  \frac{d}{dt} \left[ q\ \varphi(z)\right],
\end{array}
\end{equation}
where the first two terms in Eq. \eqref{eq:Lagrangian} describe the two oscillators. The third term in Eq. \eqref{eq:Lagrangian} is the magnetomechanical coupling rate between the motional displacement, and the small currents $\dot{q}$. The last term is a total gauge derivative, although the gauge $\varphi(z)$ is a free parameter and can be arbitrarily chosen, it is common that some specific physical conditions influence the choice of gauge. The fourth term is easily expanded as $\partial_{t} \left[ q\ \varphi(z)\right]=q \dot{z} \nabla \varphi(z)+\varphi(z) \dot{q}$, ($\partial_{t}$ is the time derivative operator) leaves the coupled equations of motion \eqref{eq:coupledequations} invariant.

The canonical flux $\phi$ and the canonical momentum $p$ are obtained through the equations  
\begin{equation}\label{eq:Momenta_Gauge}
\begin{array}{c}
\displaystyle\frac{\partial \mathcal{L }}{\partial \dot{q}}=\phi=L \dot{q} +  z\ G(z)+\varphi(z),
\\
\\ 
\displaystyle\frac{\partial \mathcal{L }}{\partial \dot{z}}=p=m \dot{z}+q\ \nabla \varphi(z).
\end{array}
\end{equation}
As we observe, the canonical momentum $p$ is a gauge dependent quantity, in our very particular case we chose $\varphi(z)=-G_0\ z$, that simplifies the ultimate form of the Hamiltonian plus recover the external capacitance in the readout circuit $C_k=m/ G_0^2$ due to the coupling rate $G_0$ \cite{caves1980measurement}. Applying the Legendre transformation $\mathscr{H}(z,q,p,\phi)=\dot{z} p +\dot{q} \phi -\mathcal{L}$ to Eq. \eqref{eq:Lagrangian}, one obtains the canonical momentum and canonical flux of the oscillators 
\begin{equation}\label{eq:generalizedmomenta}
p=m \dot{z}- G_0\  q,\hspace{0.8cm} 
\phi=L \dot{q} + z\ G(z).
\end{equation}
The canonical momentum $p$ of the coupled system includes the kinetic momentum $m \dot{z}$ and the momentum in the field $-G_0 q$. The canonical flux $\phi$ involves the current $L \dot{q}$ and an induction term $G_1\ z$. Therefore, the total classical Hamiltonian is derived from the Lagrangian through the Legendre transformation and it is given by 
\begin{equation}\label{eq:ClassHamiltonian}
\begin{array}{lll}
\mathscr{H}
=
\displaystyle\left(
\frac{p^2}{2m}
+
\omega_{\textnormal{m}}^2\frac{ m z^2}{2}
\right)&+&
\displaystyle
\left[\frac{\phi^2}{2L}
+
\left(\omega_{\textnormal{e}}^2+\frac{G_0^2}{m L}\right)\frac{ L  q^2}{2}\right]
\\
\\
&+&
\displaystyle\left( \frac{G_0}{m}\  p q
+
\frac{G_1}{L}\phi z^2\right).
\end{array}
\end{equation}
We have so far found the Hamiltonian \eqref{eq:ClassHamiltonian}, in order to quantize it, it is required to analyse the magnetomechanical single photon-phonon interaction. We define the effective linear coupling $g_0$ in terms of the zero point fluctuation of the electric charge $q_{\textnormal{\tiny ZPF}}=\sqrt{\hbar/(2 L \omega_{\textnormal{e}})}$ and mechanical momentum $p_{\textnormal{\tiny ZPF}}=\sqrt{\hbar  \omega_{\textnormal{m}}  m/2}$ such that 
\begin{equation}
\frac{\hbar g_0}{2}\equiv G_0 \ q_{\textnormal{\tiny ZPF}} \frac{p_{\textnormal{\tiny ZPF}}}{m}  ,\hspace{1cm}\frac{\hbar g_1}{2}\equiv G_1\  z_{\textnormal{\tiny ZPF}}^2\frac{\phi_{\textnormal{\tiny ZPF}}}{L}\  ,
\end{equation}
where we also defined the non-linear coupling $g_1$ in terms of the zero point motion $z_{\textnormal{\tiny ZPF}}=\sqrt{\hbar/(2 m \omega_{\textnormal{m}})}$ and zero point fluctuation of the electrical flux $\phi_{\textnormal{\tiny ZPF}}=\sqrt{\hbar L \omega_{\textnormal{e}}/2}$. Within this paper, we mostly study the regime near to resonance in which the single photon-phonon effective linear coupling is simplified as
\begin{equation}
g_0=\frac{G_0}{\sqrt{L m}}.
\end{equation}


The linear coupling is commonly characterized using spectroscopic techniques, which we discuss in section \ref{spectral}. The geometrical dependence of the  linear coupling $g_0$ is described in \ref{ExperimentalConsiderations}.

\section{Quantum dynamics}\label{QuantumDynamics}

In this section we explore the quantum magnetomechanical Hamiltonian and some potential applications of a quantum system of this physical characteristics. Following the standard process in opto- and electro-mechanics \cite{law1995Interaction}, we quantize the classical Hamiltonian \eqref{eq:ClassHamiltonian} with the standard commutation relations $[\hat{q},\hat{p}]=[\hat{z},\hat{\phi}]=[\hat{z},\hat{q}]=[\hat{p},\hat{\phi}]=0,$ and $[\hat{q},\hat{\phi}]=[\hat{z},\hat{p}]=i\hbar$. The quantum magnetomechanical Hamiltonian $\hat{H}_{m}$ is given by
\small
\begin{equation}\label{eq:magmechanicH}
\begin{array}{lcr}
\hat{H}_{m}&=& 
\left(
\displaystyle\frac{\hat{\phi}^2}{2L}
+
(\omega_{\textnormal{e}}^2 +g_0^2)\frac{L \hat{q}^2}{2}
\right)+
\left(\displaystyle\frac{\hat{p}^2}{2m}
+
\Omega_{\textnormal{m}}(\hat{\phi})^2 \frac{m\hat{z}^2}{2}
\right)\\
\\
& &+
g_0
\sqrt{\frac{L}{m}}\hat{p}\hat{q},
\end{array}
\end{equation}
\normalsize
where the mechanical frequency is modulated by the flux in the $LC$ circuit as
\begin{equation}\label{eq:OmegaNL}
\Omega_{\textnormal{m}}^2(\hat{\phi})
=
\omega_{\textnormal{m}}^2-\frac{2 g_1}{\sqrt{L m}}\hat{\phi}.
\end{equation}
The Lorentz force exerted between the permanent magnet and the field generated by the current induces the modulation of the mechanical frequency $\Omega_{\textnormal{m}}$. The effect is known as Lorentz spring constant $k_L=-2 g_1\sqrt{m/L}$ as a result of the modification of the total stiffness $k=k_0+k_L$ of the mechanical oscillator.

The magnetomechanical Hamiltonian clearly allows to perform mechanical frequency modulation through the Lorentz force 
\begin{equation}
\hat{F}_{L}=-\frac{\partial \hat{H}_{int}}{\partial \hat{z}}=2 g_1 \sqrt{\frac{m}{L}}\hat{z} \hat{\phi}.
\end{equation}
One of the applications of the Lorentz force at the mesoscale is the implementation of it's back action to cool down the mechanical motion of mechanical oscillators \cite{wang2008cooling}. The non-linear properties of the Hamiltonian in Eq. \eqref{eq:magmechanicH} represent a novel introduction for the non-linear dynamics of mechanical systems \cite{sankey2010strong}. The second order non-linear interaction $(\hat{\phi} \hat{z}^2)$ of Eq. \eqref{eq:Hinteraction} induces an $x$-squared type non-linearity allowing to produce mechanical squeezing \cite{szorkovszky2014detuned}, mechanical amplification \cite{levitan2016optomechanics}, mechanical entanglement \citep{brawley2016nonlinear} or cooling through mechanical frequency modulation \cite{bienert2015optomechanical}.

In the general magnetomechanical interaction, we look at two different regimes of interest depending on the strength of the coupling rates $g_0$ and $g_1$. Since in general $|g_0|\gg |g_1|$, the terms involving $g_1^2$ are usually negligible, we may write the magnetomechanical Hamiltonian as a sum of linear and non-linear terms, i.e. $\hat{H}_{m}=\hat{H}_{L}+\hat{H}_{NL}$, where   
\begin{equation}\label{eq:HamiltonianLinear}
\hat{H}_L=\frac{\hat{\phi}^2}{2L}
+
(\omega_{\textnormal{e}}^2+g_0^2)\frac{ L  \hat{q}^2}{2}
+ 
\frac{\hat{p}^2}{2m}
+
\omega_{\textnormal{m}}^2\frac{ m \hat{z}^2}{2}
+
g_0
\sqrt{\frac{L}{m}}\hat{p}\hat{q},
\end{equation}
and 
\begin{equation}\label{eq:Hinteraction}
\hat{H}_{NL}
\approx\-g_1 \displaystyle \sqrt{\frac{m}{L}}\hat{z}^2\hat{\phi}.
\end{equation}
With $H_L$ alone we recover a scheme proposed for quantum non-demolition measurements and velocity sensing \cite{caves1980measurement}.


In the optomechanics community, the Hamiltonian is typically expressed in the boson representation. In order to facilitate the comparison between magnetomechanics and optomechanics, here we re-express our magnetomechanical Hamiltonian \eqref{eq:magmechanicH} in the boson operators representation
\small
\begin{equation}\label{eq:magmechanicHBoson}
\begin{array}{lcr}
\hat{H}_{m}&=& \hbar \omega_e \hat{a}^{\dagger}\hat{a}
+
\hbar \Omega_{\textnormal{m}}(\hat{\phi}) \hat{b}^{\dagger}\hat{b} + i\displaystyle\frac{\hbar g_0}{2}\left(\hat{a}^{\dagger}+\hat{a}\right)\left(\hat{b}-\hat{b}^{\dagger}\right)\\
\\
& & 
+
\displaystyle\frac{\hbar g_0^2}{4 \omega_e}(\hat{a}+\hat{a}^{\dagger})^2
\end{array}
\end{equation}
\normalsize 
We introduce the boson creation $\hat{a}^\dagger$ ($\hat{b}^\dagger$) and the annihilation $\hat{a}$ ($\hat{b}$) operators for the electromagnetic (acoustic) field. The boson operators are defined by the relations $\hat{q}=q_{\textnormal{\tiny ZPF}}(\hat{a}+\hat{a}^{\dagger})$, $\hat{\phi}=i \phi_{\textnormal{\tiny ZPF}}(\hat{a}^{\dagger}-\hat{a})$, $\hat{z}=z_{\textnormal{\tiny ZPF}}(\hat{b}+\hat{b}^{\dagger})$ and $\hat{p}=i p_{\textnormal{\tiny ZPF}}(\hat{b}^{\dagger}-\hat{b})$. The boson operators act on the eigenstates of the electromagnetic (acoustic) field mode $|n_{\textnormal{e}} \rangle$ ($|n_{\textnormal{m}} \rangle$) following the standard raising $\hat{a}^{\dagger}|n_{\textnormal{e}} \rangle =\sqrt{n_{\textnormal{e}} +1}|n_{\textnormal{e}} +1\rangle$ ($\hat{b}^{\dagger}|n_{\textnormal{m}} \rangle =\sqrt{n_{\textnormal{m}} +1}|n_{\textnormal{m}} +1\rangle$) and lowering $\hat{a}|n_{\textnormal{e}} \rangle =\sqrt{n_{\textnormal{e}} -1}|n_{\textnormal{e}} -1\rangle$ ($\hat{b}|n_b\rangle =\sqrt{n_{\textnormal{m}} -1}|n_{\textnormal{m}} -1\rangle$) relations. The eigenvector basis for the magnetomechanical states is described by $|n_{\textnormal{e}} ,n_{\textnormal{m}} \rangle=|n_{\textnormal{e}} \rangle \otimes |n_{\textnormal{m}} \rangle$.

\section{Linear Quantum Magnetomechanics}\label{LinearRegime}
In this section, we focus our study on the linear magnetomechanical Hamiltonian $\hat{H}_L$, where $g_0\gg g_1$. This raises a dominant linear interaction defined by the charge-momentum coupling $\hat{q} \hat{p}$. In regular optomechanics, the interaction is commonly described by a linearized model with a bi-linear position-position coupling \cite{aspelmeyer2014cavity}. In the magnetomechanical linear interaction picture we explore two different regimes the so- called strong coupling regime and ultra-strong coupling regime. Charge-momentum coupling remains little explored, and to the best of our knowledge there are no proposals demonstrating that ultra-strong coupling for mechanical systems can be achieved in this fashion. We also note that our magnetomechanical system breaks the time reversal symmetry (Appendix A).

\subsection{Strong coupling regime}\label{SC}

In the magnetomechanical strong coupling regime the interaction between the mechanics and the electronics is faster than the decoherence for each individual resonator $g_0^{-1}< \Gamma_{\textnormal{m}}^{-1},\Gamma_e^{-1}$. In the strong coupling regime $g_0\ll \omega_{\textnormal{m}}, \omega_{\textnormal{e}}$ and the term $g_0^2/ \omega_{\textnormal{e}} \ll g_0$. The elements in \eqref{eq:HamiltonianLinear} in the boson basis with terms proportional to $g_0^2/\omega_{\textnormal{e}}$ are negligible and the simplified Hamiltonian in the strong coupling regime is 

 

\begin{equation}\label{eq:magmechanicHBoson}
\hat{H}_{\textnormal{\tiny SC}}=\hbar \omega_{\textnormal{e}} \hat{a}^{\dagger} \hat{a}+\hbar \omega_{\textnormal{m}} \hat{b}^{\dagger}\hat{b}
 + i\frac{\hbar g_0}{2}\left(\hat{a}^{\dagger}+\hat{a}\right)\left(\hat{b}-\hat{b}^{\dagger}\right)
\end{equation}
which is easily diagonalized as the sum of two normal modes $\hat{H}_{\textnormal{\tiny SC}}=\hbar \omega_{\textnormal{\tiny SC}+}\hat{c}_{+}^{\dagger}\hat{c}_{+}+\hbar \omega_{\textnormal{\tiny SC}-}\hat{c}_{-}^{\dagger}\hat{c}_{-}$. The normal modes $\hat{c}_{\pm}$ are a hybridized mode that contains phonon and photon modes. The energy levels for the hybrid system are  
\begin{equation}\label{eq:EnergySC}
E_{\textnormal{\tiny SC}\pm}=\frac{\hbar}{\sqrt{2}}
\left(
\omega_{\textnormal{m}}^2+\omega_{\textnormal{e}}^2
\pm 
\sqrt{
4 g_0^2 \omega_{\textnormal{e}}^2
+
\left(\omega_{\textnormal{m}}^2-\omega_{\textnormal{e}}^2\right)^2
}
\right)^{1/2}
\end{equation}

\begin{figure}[h]
\includegraphics[scale=0.29]{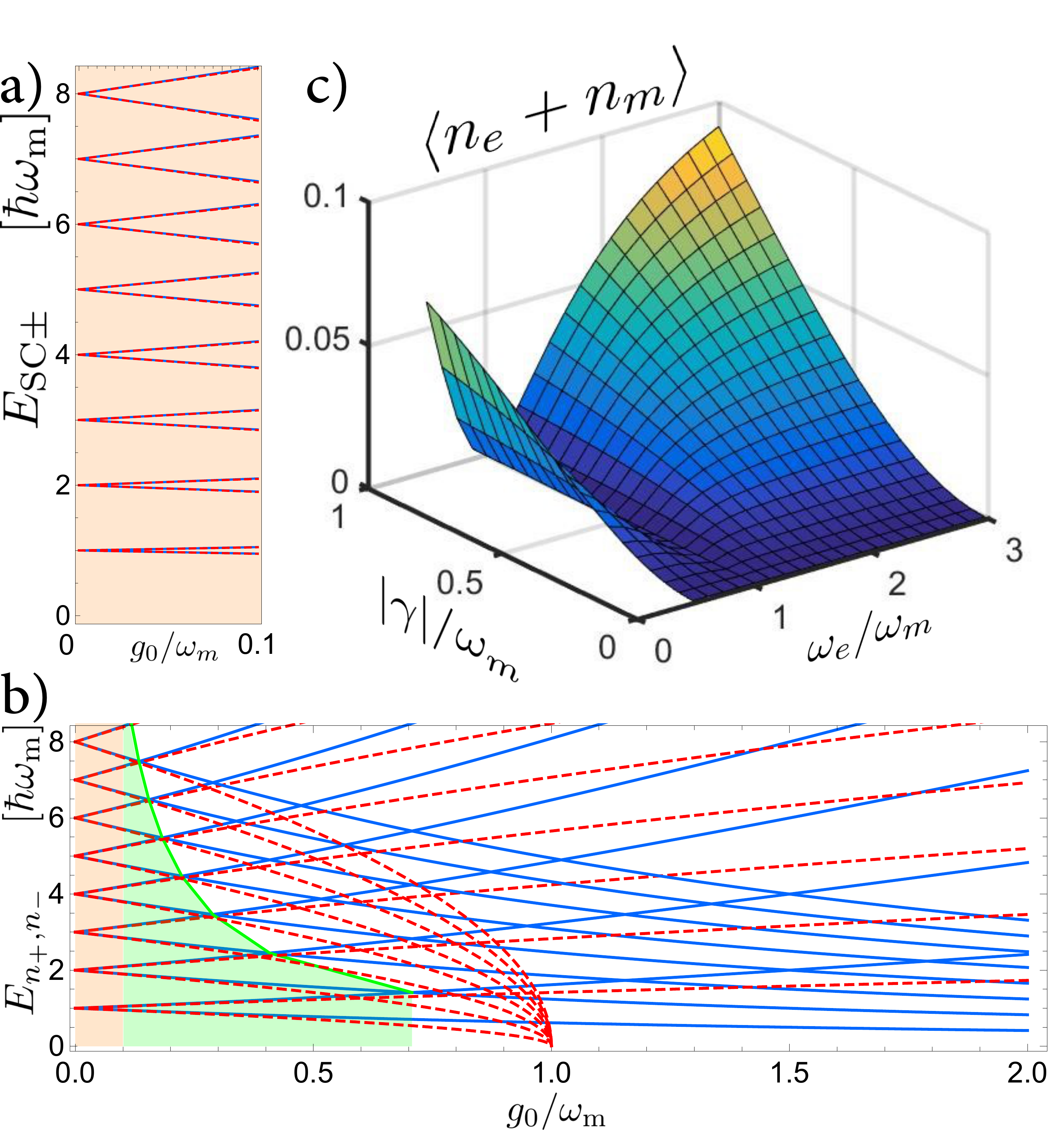}
\caption{In blue lines, the generalized spectrum and in red dashed lines the approximation for the spectra in the strong coupling regime. a) Energy spectra of the magnetomechanical system within the strong coupling regime $g_0/\omega_{\textnormal{m}}$.  b) Energy spectra of the magnetomechanical system in two different perturbative regimes, strong coupling (orange background $0\leq g_0/\omega_{\textnormal{m}}\leq 0.1$) and ultra-strong coupling (green background $0.1\leq g_0/\omega_{\textnormal{m}}\leq 1/\sqrt{2}$). The green line represents the Juddian points that delimit the perturbative ultra-strong coupling regime. The spectra a) and b) were calculated considering in resonance a resonant system $\omega_{\textnormal{e}}=\omega_{\textnormal{m}}$. c) Expectation value of the total excitation number populating the ground state $\langle n_{\textnormal{e}} +n_{\textnormal{m}} \rangle $. The calculation was performed using a truncation of 15 Fock states. 
}\label{fig:GroundState}
\end{figure}

The spectrum for the first eight eigenvalues are shown in red dashed lines in Fig. \ref{fig:GroundState}a as a function of $g_0/\omega_{\textnormal{m}}$ for values that lay within the strong coupling regime ($\Gamma_{\textnormal{m}}/\omega_{\textnormal{m}}<g_0/\omega_{\textnormal{m}}\leq0.1 $), in blue the values obtained for the general solution discussed in the next section. In this regime we observe a very typical linear dependence and good agreement between the general and strong coupling approximation.

The Hamiltonian Eq. \eqref{eq:magmechanicHBoson} shows that in the strong coupling regime, the magnetomechanical linear system allows to perform linear operations available in optomechanics such as state swap between the mechanics and the electronics, cooling or heating of the mechanical oscillator through a magnetomechanical protocol, squeezing of the mechanical mode or implementation of \textit{quantum non-demolition} protocols. The terms $\propto (\hat{a}^{\dagger}\hat{b}-\hat{a}\hat{b}^{\dagger})$ in the Hamiltonian Eq. \eqref{eq:magmechanicHBoson} represent the energy exchange between electronic and mechanical mode, commonly known as beam splitter interaction and crucial for state transfer protocols. Meanwhile the terms $\propto (\hat{a}\hat{b}-\hat{a}^{\dagger}\hat{b}^{\dagger})$ are simultaneous excitations of the mechanical and electromagnetic field, known as two mode squeezing interaction \cite{aspelmeyer2010quantum,felicetti2014dynamical}. The magnetomechanical system in the linear regime reveals a novel interface to implement hybrid mechanical systems with strong interactions. 
 
In section \ref{LangevinEquations} we expand the discussion of spectral properties and and suitable measurements for the strong magnetomechanical coupling regime.


\subsection{Perturbative ultra-strong coupling regime}\label{USC}

The magnetomechanical system offers a new platform for the exploration of regimes beyond the strong coupling, where relevant phenomena have been described in modern literature \cite{rossatto2016entangling,
bourassa2009ultrastrong,
casanova2010deep}. In the ultra strong coupling regime that has been predominantly explored with electronic circuits \cite{forn2010observation,niemczyk2010circuit,bourassa2009ultrastrong}, the magnetomechanical interface opens the possibility for mechanical systems to reach this developing regime. In order to identify the different regimes, spectral properties have been identified \cite{rossatto2017spectral,hausinger2010qubit}. The treatment for $\hat{H}_L$ in the ultra-strong coupling becomes a more complicated task when terms in the Hamiltonian are not negligible any more. Here we consider a lossless environment $\Gamma_{\textnormal{m,e}}= 0$ and find a general diagonalization for $\hat{H}_L$ and the construction of the eigenstates of the magnetomechanical system and spectral properties.

\subsubsection{Eigenstates of the magnetomechanical system}\label{Eigenstates_Magnetomechanics}

To construct the eigenstates of the magnetomechanical system beyond the strong coupling approximation is needed to diagonalize the Hamiltonian $\hat{H}_L$ into its normal modes. The normal modes are given by $\hat{H}_{\textnormal{\tiny NM}}=\hat{U}\hat{H}_{L}\hat{U}^{\dagger}$ which is diagonal and has normal modes frequencies $\omega_{\pm}$. The unitary transformation $\hat{U}$ is a two mode squeezing operator $\hat{U}=
\exp \left\{
i \beta 
\left(e^{-\gamma} \hat{p} \hat{\phi}
-
e^{\gamma} \hat{z} \hat{q}
\right)\right\}$ following \cite{hong2007one} with complex squeezing parameter is $e^{\gamma}=i\ L m\ \omega_{\textnormal{m}} \sqrt{\omega_{\textnormal{m}} \omega_{\textnormal{e}}}$. The diagonalized linear lossless Hamiltonian $\hat{H}_{\textnormal{\tiny NM}}$ is expressed in the normal mode basis as
\begin{equation}\label{eq:Hamiltoniandiagonal}
\hat{H}_{\textnormal{\tiny NM}}
=
\frac{\omega_-^2}{2}\left(\hat{X}_-^2+\hat{P}_-^2\right)
+
\frac{\omega_+^2}{2}\left(\hat{X}_+^2
+
\hat{P}_+^2\right),
\end{equation} 
where the eigenfrequencies are given by
\begin{equation}\label{eq:eigenfrequencies}
\begin{array}{lll}
\omega_{\pm}^2&=&\frac{1}{2}
\left(
\omega_{\textnormal{m}}^2 + \omega_{\textnormal{e}}^2+g_0^2\right)
\\
\\
&\pm& 
\frac{1}{4}\sqrt{
4g_0^2 (\omega_{\textnormal{e}}^2+g_0^2)
+
\left(\omega_{\textnormal{m}}^2-\omega_{\textnormal{e}}^2-g_0^2\right)^2
}
.
\end{array}
\end{equation}

The dimensionless quadratures in the normal mode basis for position $\hat{X}_{\pm}$ and momentum $\hat{P}_{\pm}$ follow the standard commutation relations $[\hat{X}_{\pm},\hat{P}_{\pm}]=i$ and each one of the quadratures is defined as
\begin{equation}
\begin{array}{lc}
\hat{X}_+=
\displaystyle\frac{1}{\sqrt{2} q_{\textnormal{\tiny ZPF}}}\left(\hat{q} \cosh{\beta}+e^{-\gamma}\hat{p}\sinh{\beta}\right),&\\
\\
\hat{P}_+=\displaystyle\frac{1}{\sqrt{2} \phi_{\textnormal{\tiny ZPF}}}(\hat{\phi} \cosh{\beta}+ e^{\gamma} \hat{z}\sinh{\beta}),&\\
\\
\hat{X}_-=\displaystyle\frac{1}{\sqrt{2}  z_{\textnormal{\tiny ZPF}}}(\hat{z}\cosh{\beta}+e^{-\gamma}\hat{\phi} \sinh{\beta}),&\\
\\
\hat{P}_-=\displaystyle\frac{1}{\sqrt{2}  p_{\textnormal{\tiny ZPF}}}(\hat{p}\cosh{\beta}+e^{\gamma} \hat{q}\sinh{\beta}),&
\end{array}
\end{equation}
where $i \tanh{(2 \beta) }=\frac{2 g_0 \omega_{\textnormal{m}}}{g_0^2 + \omega_{\textnormal{e}}^2 - \omega_{\textnormal{m}}^2}$.
 
As a matter of completeness, we introduce the boson creation and annihilation operators for the $\pm$ modes, defined as 
\begin{equation}\label{eq:bosonoperators}
\begin{array}{c}
\hat{a}_+=\displaystyle\frac{1}{\sqrt{2}}\left(\hat{X}_+ +i \hat{P}_+ \right),\hspace{0.5cm}
\hat{a}_+^{\dagger}=\displaystyle\frac{1}{\sqrt{2}}\left(\hat{X}_+ -i \hat{P}_+ \right),\\
\\
\hat{a}_-=\displaystyle\frac{1}{\sqrt{2}}\left(\hat{X}_- +i \hat{P}_- \right),\hspace{0.5cm}
\hat{a}_-^{\dagger}=\displaystyle\frac{1}{\sqrt{2}}\left(\hat{X}_- -i \hat{P}_- \right).
\end{array}
\end{equation}
With the boson operators Eq. \eqref{eq:bosonoperators} defined, it is straightforward to determine the eigenstates of the magnetomechanical system in the linear regime, which are
\begin{equation}\label{eq:eigenstates}
|n_+,n_-\rangle =\frac{1}{\sqrt{n_+ ! n_-!}}(\hat{a}_+^{\dagger})^{n_+}(\hat{a}_-^{\dagger})^{n_-}|0,0\rangle,
\end{equation}
with the raising operators $\hat{a}_{\pm}$ acting on the vacuum state $|0,0\rangle$. Some of the relevant properties of the boson operators Eq. \eqref{eq:bosonoperators} are the standard commutation relations $[\hat{a}_+,\hat{a}_+^{\dagger}]=1$ and $[\hat{a}_-,\hat{a}_-^{\dagger}]=1$. Similarly the number operator for the bosonic modes are $\hat{n}_+\equiv \hat{a}_+^{\dagger}\hat{a}_+$ and $\hat{n}_-\equiv \hat{a}_-^{\dagger}\hat{a}_-$, with expectation values   $n_{\pm}=\langle\hat{n}_{\pm}\rangle$. Once the system has been expressed in the boson operator representation, it is clear that neglecting the vacuum energy, the system has the following energy spectrum
\begin{equation}\label{eq:eigenvalues}
E_{n_+,n_-}=\hbar \omega_+ n_+ +\hbar \omega_- n_- .
\end{equation}

The energy spectrum Eq. \eqref{eq:eigenvalues} as a function of $g_0/\omega_{\textnormal{m}}$ with its first eight eigenvalues is shown in Fig. \ref{fig:GroundState}b in continuum blue lines. In red dashed lines its shown the spectrum Eq. \eqref{eq:EnergySC} which corresponds to the strong coupling regime approximation. It is clear that for small values of $0 \leq g_0/\omega_{\textnormal{m}} \leq 0.1$ the strong coupling spectrum of Fig. \ref{fig:GroundState}a accurately describes the energy levels of the system as the red dashed lines mostly overlap the blue lines. A modern quantitative definition for classification of coupling regimes according to spectral properties \citep{rossatto2017spectral} suggest that the so-called ultra strong coupling regime can be separated into perturbative and non-perturbative ultra strong coupling regime. Where the perturbative ultra strong coupling regime is defined as the region where $g_0/\omega_{\textnormal{m}}\leq J_n$ where $J_n$ are the first Juddian points of the spectra. We calculated the Juddian points for the spectra of the magnetomechanical system 
\begin{equation}
J_n=\frac{1}{\sqrt{n_-+n_-^2}},
\end{equation}
are shown in \ref{fig:GroundState}b as black crosses. According to Rossatto et. al. \citep{rossatto2017spectral}, the perturbative ultra strong coupling regime will be delimited by the first Juddian point for $n_-=1$. The magnetomechanical perturbative ultra-strong coupling regime is then defined for coupling within $g_0/\omega_{\textnormal{m}}\leq \frac{1}{\sqrt{2}}$. The coupling rate $g_0$ can be experimentally measured directly from the electrical resonance frequency shift $\omega_{\textnormal{e}}\rightarrow\Omega_{\textnormal{e}}$, with $\Omega_e=\sqrt{\omega_{\textnormal{e}}^2+g_0^2}$. If the value of the frequency shift is negligible, the energy spectrum \eqref{eq:EnergySC} defines the energy levels.

\subsubsection{Entangled Ground State}\label{GS}

As we defined the diagonalized Hamiltonian in the normal modes basis Eq. \eqref{eq:Hamiltoniandiagonal}, when they act on a on an eigenstate $|n_+ ,n_-\rangle$, we obtain  $\hat{H}_{\textnormal{NM}}|n_+ ,n_- \rangle=E_{n_+,n_-}|n_+,n_- \rangle$. We can transform these eigenstates back to the lab frame using the unitary $\hat{U}^\dagger$, and denote them as $\overline{|n_+,n_-  \rangle}
\equiv \hat{U}^{\dagger}|n_+,n_-\rangle$. We note that in the lab basis such eigenstates may be entangled. To see this we look at the expectation values for the standard occupations for the number operators for the electric and mechanical excitations $\hat{n}_e=\hat{a}^\dagger\hat{a},\;\; \hat{n}_m=\hat{b}^\dagger\hat{b}$, in this lab frame. In particular we compute the sum, $\langle\hat{n}_{\textnormal{m}} +\hat{n}_{\textnormal{e}} \rangle\equiv\overline{\langle 0,0|} \hat{n}_e\overline{|0,0\rangle}+\overline{\langle 0,0|} \hat{n}_m\overline{|0,0\rangle}$, (with a Fock truncation of 15), and in Fig.\ref{fig:GroundState}c we plot this sum as a function of $\omega_e/\omega_m$ and $|\gamma|/\omega_m$.
We see that when $|\gamma|=0$, i.e. when there is no coupling between the electric and mechanical systems, the ground state has no excitations. However this is no longer true when $|\gamma|>0$ and $\omega_e\ne\omega_m$.
By squeezing the zero-point fluctuations of the magnetomechanical system the ground state becomes entangled. Ground-state entanglement induces the emergence of negative energy-density regions in quantum systems \cite{hotta2009ground}. The entanglement present when $\langle\hat{n}_{\textnormal{m}} +\hat{n}_{\textnormal{e}} \rangle> 0$ represents a signature of the quantum nature of the magnetomechanical system. This magnetomechanical system presents a novel approach for the generation of negative energy-density which can be implemented in protocols of quantum energy teleportation \cite{hotta2009ground}.


\section{Nonequilibrium system}\label{LangevinEquations}

So far, we have described the magnetomechanical system in an isolated environment i.e. in the absence of decoherence. In this section we consider a semi-classical description in the presence of decoherence channels $\Gamma_e,\Gamma_{\textnormal{m}}\neq 0$ and the response of the mechanical system to thermal excitations. In section \ref{susceptibility} we describe observable properties such as magnetomechanical damping and magnetomechanical frequency shift. 


The simplest dynamics of the magnetomechanical system out of equilibrium arises when we consider the mechanical system to be in contact with a thermal bath through the decoherence channel $\Gamma_{\textnormal{m}}\neq 0$. In this case we consider the external driving force $F_{\textnormal{ext}} (t)$ no longer dominated by the Lorentz force $F_{L}(t)$ but by the random thermal Langevin force $F_{\textnormal{th}} (t)$. The mechanical system is then driven by $F_{\textnormal{th}} (t)$ and as a result it has a randomly time-varying amplitude and phase. In most experiments, the oscillations of micro scale mechanical systems are analysed as a noise spectrum in frequency space. Here we describe the stationary spectral properties of the magnetomechanical system and analyse the influence of the $LC$ circuit on the mechanics. 

The fluctuations of the mechanical displacement are a consequence of Brownian motion due to the fact that the mechanical oscillator is driven by a noisy thermal force. We describe the system using the Langevin equation $\partial \hat{\mathcal{O}}/\partial t = (i/\hbar) [\hat{H}_L,\hat{\mathcal{O}}]+\hat{\mathcal{N}}_{\mathcal{O}}$ for an arbitrary observable $\hat{\mathcal{O}}$, where $\hat{\mathcal{N}}_{\mathcal{O}}$ represents the noise introduced by the interaction of the observable $\hat{\mathcal{O}}$ with it's environment. We calculate the Langevin equation for the coupled system of observables $\hat{z}, \hat{p},\hat{q}$ and $\hat{\phi}$ which reads as 
\begin{equation}\label{eq:Langevin}
\begin{array}{ll}
\dot{\hat{z}}=& \displaystyle\frac{\hat{p}}{m}+g_0\sqrt{\frac{L}{m}}\hat{q}\\
\\
\dot{\hat{p}}=& m \omega_{\textnormal{m}}^2 \hat{z}-\Gamma_{\textnormal{m}} \hat{p} -\hat{F}_{\textnormal{ext}} (t)\\
\\
\dot{\hat{q}}=& \displaystyle\frac{\hat{\phi}}{L}\\
\\
\dot{\hat{\phi}}=&  L \Omega_e^2 \hat{q}-\Gamma_{\textnormal{e}}\hat{\phi} -g_0\displaystyle\sqrt{\frac{L}{m}}\hat{p} +\hat{V}_{\textnormal{ext}} (t).
\end{array}
\end{equation}
It is clear that the noisy elements are introduced in the "momentum" terms as they are commonly associated to friction forces. From the experimental point of view, it is easier to measure the properties of the system in the frequency domain, looking at the stationary case. For the stationary case its possible to consider that we measure continuously for a finite time $\tau$, in this situation, the frequency components of the displacement is (and the definition is extended to all the other operators)
\begin{equation}
\tilde{z}(\omega)=\frac{1}{\sqrt{\tau}}\int_0^{\tau} \hat{z}(t) e^{i \omega t }dt.
\end{equation}
For the limit $\tau\rightarrow \infty$, the response of the mechanical oscillator to an external drive is $\tilde{z}(\omega)=\chi_{\textnormal{m}}(\omega)\tilde{F}_{\textnormal{ext}} (\omega)$, where the susceptibility of the mechanics is $\chi_{\textnormal{m}}(\omega)=(m(\omega_{\textnormal{m}}^2-\omega^2+i \omega \Gamma_{\textnormal{m}}))^{-1}$.

In the regime where $g_0>0$, we calculate the expected value of the operators that we obtained from the Langevin equation and transformed into the frequency domain $\langle\tilde{z}\rangle$, $\langle\tilde{p}\rangle$, $\langle\tilde{q}\rangle$ and $\langle\tilde{\phi}\rangle$. One obtains a set of coupled equations represented in matrix form as ${\bf Y}=({\bf R}+ i \omega {\bf I}){\bf Y}$ which has normal mode frequencies, 
\begin{equation}\label{eq:eigenfrequenciesDamped}
\Omega_{\pm}^2=\frac{1}{2}\Xi^2
\pm
\frac{1}{2} 
\sqrt{4 g_0^2 \omega_{\textnormal{m}}^2
+\Xi^4
-
4 \omega_{\textnormal{m}}^2\Omega_e^2},
\end{equation} 
where $\Xi^2=\omega_{\textnormal{m}}^2+\Omega_e^2 +\Gamma_{\textnormal{e}} \Gamma_{\textnormal{m}} $, the vector ${\bf Y}=\left(\langle \tilde{z}\rangle,\langle \tilde{p}\rangle,\langle\tilde{q}\rangle,\langle \tilde{\phi}\rangle\right)$, and
\begin{equation}\label{eq:CoupledFrequency}
{\bf R}=
\left(
\begin{array}{cccc}
0 &\frac{1}{m}& \sqrt{\frac{L}{m}} g_0 & 0 \\
-m \omega_{\textnormal{m}}^2 & -\Gamma_{\textnormal{m}} & 0 & 0 \\
0 & 0 & 0 & \frac{1}{L} \\
0&-\sqrt{\frac{L}{m}} g_0 & -L \Omega_e^2 &-\Gamma_{\textnormal{e}} \\
\end{array}
\right).
\end{equation}

The normal modes frequencies are shown in Fig. \ref{fig:NMS}a and \ref{fig:NMS}b as red dashed lines. The coupling $g_0$ is characterized experimentally by the splitting $\Omega_+-\Omega_-\approx g_0$ in the power spectral density (PSD) \cite{groblacher2009observation}. If the splitting is observable, it is a signature of strong coupling between the modes. Below we find that strong coupling can be achieved in the linear regime of this magnetomechanical system.

\subsection{Mechanical Susceptibility}\label{susceptibility}

In a self contained fashion, the presence of the magnetomechanical coupling $g_0>0$ creates a "circulation" of energy. The mechanical displacement generates a voltage in the $LC$ resonator while this one generates a magnetic field that exerts a force on the mechanical resonator. Intuitively, it is clear that the coupling might modify the bare mechanical susceptibility $\chi(\omega)$. This modified mechanical susceptibility is now called the effective susceptibility $\chi_{\text{eff}}(\omega)$ and is obtained from the solution for the set of coupled equations $\bf Y$. We can express $\chi_{\text{eff}}(\omega)$ in terms of the mechanical bare susceptibility plus a magnetomechanical modification $\Sigma(\omega)$ such that 
\begin{equation}
\chi_{\text{eff}}(\omega)=\frac{1}{m(\omega_{\textnormal{m}}^2-\omega^2+i \omega \Gamma_{\textnormal{m}})+\Sigma(\omega)},
\end{equation} 
where similarly to optomechanics, $\Sigma(\omega)$ represents the so called self-energy \cite{aspelmeyer2014cavity}. The modification of the mechanical susceptibility can be classified into a magnetomechanical induced damping rate $\Gamma_{\text{mm}}(\omega)=- \text{Im}[\Sigma(\omega)]/m \omega$ and a magnetomechanical induced frequency shift $\delta \omega_{\textnormal{m}}(\omega)=\text{Re}[\Sigma (\omega)]/2m \omega $. Explicitly, these parameters take the form of the magnetomechanical damping
\begin{equation}\label{eq:Gamma_mm}
\Gamma_{\text{mm}}(\omega)=-g_0^2
\left[
\frac{\omega^2 (\Gamma_{\textnormal{e}}-\Gamma_{\textnormal{m}})+\Gamma_{\textnormal{m}} \omega_{\textnormal{e}}^2}
{\Gamma_{\textnormal{e}}^2 \omega^2+\left(\omega^2-\omega_{\textnormal{e}}^2\right)^2}\right]
\end{equation}
and magnetomechanical induced frequency shift
\begin{equation}
\delta \omega_{\textnormal{m}}(\omega)
=\frac{
g_0^2 \omega}{2} 
\left[\frac{  \omega_{\textnormal{e}}^2-\omega^2-\Gamma_{\textnormal{e}} \Gamma_{\textnormal{m}}}
{\Gamma_{\textnormal{e}}^2 \omega^2+\left(\omega^2-\omega_{\textnormal{e}}^2\right)^2}\right].
\end{equation} 
The control of parameters such as $\Gamma_{\textnormal{mm}}(\omega)$ and $\delta \omega_{\textnormal{m}}(\omega)$ make it possible to implement protocols such as cooling or heating of the mechanical oscillator through the $LC$ resonator within this magnetomechanical approach.

\subsection{Spectral Properties}\label{spectral}

\begin{figure}[h]
\includegraphics[scale=0.6]{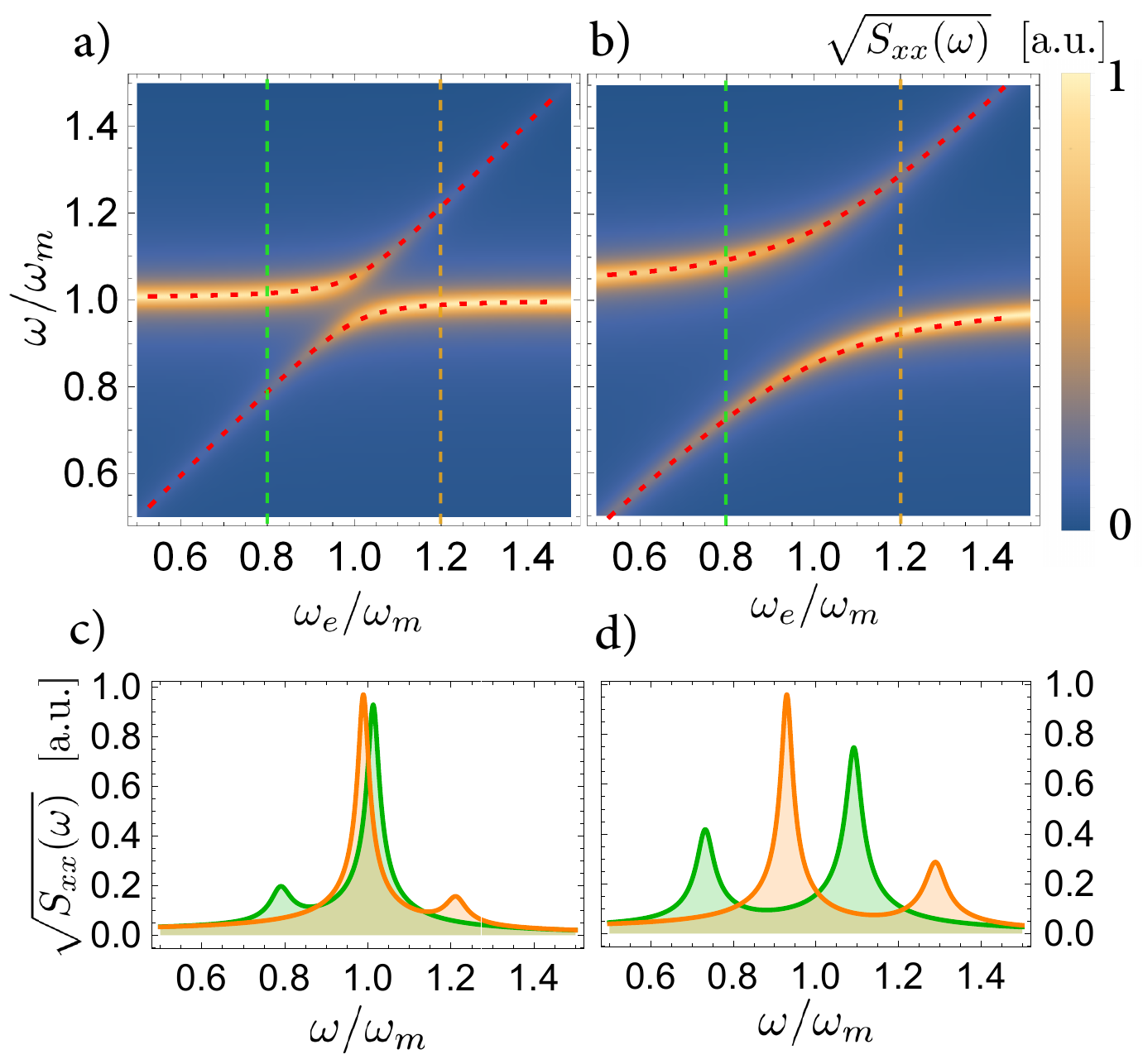}
\caption{a) Plot of the position displacement $\sqrt{\langle |z|^2\rangle}=\sqrt{S_{\text{xx}}(\omega)}$ [a.u.] considering the parameters $(\Gamma_{\textnormal{m}},\Gamma_{\textnormal{e}},g_0)=(0.025,0.05,0.1)\times\omega_{\textnormal{m}}$ as a function of the frequency $\omega/\omega_{\textnormal{m}}$ and $\omega_{\textnormal{e}}/\omega_{\textnormal{m}}$, where we set $m=L=1$. The normal modes frequencies $\Omega_{\pm}$ are shown as red dashed lines and avoided crossing is observed. For $\omega_{\textnormal{e}}=0.8\omega_{\textnormal{m}}$ (green dashed line) and $\omega_{\textnormal{e}}=1.2\omega_{\textnormal{m}}$ (orange dashed line) we show the profile in Fig. \ref{fig:NMS}c which is normalized to the maximum. b) Plot of the mechanical response considering the parameters $(\Gamma_{\textnormal{m}},\Gamma_{\textnormal{e}},g_0)=(0.025,0.05,0.3)\times\omega_{\textnormal{m}}$ as a function of the frequency $\omega/\omega_{\textnormal{m}}$ and $\omega_{\textnormal{e}}/\omega_{\textnormal{m}}$. The normal modes frequencies $\Omega_{\pm}$ are shown as red dashed lines and avoided crossing is observed. For $\omega_{\textnormal{e}}=0.8\omega_{\textnormal{m}}$ (green dashed line) and $\omega_{\textnormal{e}}=1.2\omega_{\textnormal{m}}$ (orange dashed line) we show the profile in Fig. \ref{fig:NMS}d, which is normalize to the maximum.}\label{fig:NMS}
\end{figure}

Commonly, the properties of the mechanical systems are experimentally characterized by measuring the power spectral density (PSD) which we define as  
\begin{equation}
S_{\text{xx}}(\omega)
=
\langle |\tilde{z}(\omega)|^2\rangle
=\langle |\chi_{\text{eff}}(\omega)|^2\tilde{F}_{\textnormal{th}} (\omega)\rangle.
\end{equation}
The PSD has units of $\text{m}^2/\text{Hz}$ and represents the distribution of energy in each frequency component of the signal \cite{bowen2015quantum}. In the case where the thermal energy drives the mechanical oscillation, it is possible to relate the variance of the amplitude of the oscillation to the thermal energy stored in the oscillator by the \textit{fluctuation dissipation theorem} $\langle|\tilde{z}(\omega_{\textnormal{m}})|^2\rangle=k_B\ T\ \Gamma_{\text{\text{eff}}}(m \omega_{\textnormal{m}})^{-1}$, where $\Gamma_{\text{eff}}=\Gamma_{\textnormal{m}}+\Gamma_{\text{mm}}$. The amplitude of the oscillation is then related as $\sqrt{S_{\text{xx}}(\omega)}=\sqrt{\langle |\tilde{z}(\omega)|^2\rangle}$. In Fig. \ref{fig:NMS}a and Fig. \ref{fig:NMS}c we plot $\sqrt{S_{\text{xx}}(\omega/\omega_{\textnormal{m}})}$ with values $(\Gamma_{\textnormal{m}},\Gamma_{\textnormal{e}},g)=(0.025\omega_{\textnormal{m}},0.05\omega_{\textnormal{m}},0.1\omega_{\textnormal{m}})$. In Fig. \ref{fig:NMS}b and \ref{fig:NMS}d the considered values are $(\Gamma_{\textnormal{m}},\Gamma_{\textnormal{e}},g)=(0.025\omega_{\textnormal{m}},0.05\omega_{\textnormal{m}},0.3\omega_{\textnormal{m}})$. The Fig. \ref{fig:NMS}c and \ref{fig:NMS}d show two plots for $\tilde{z}(\omega)$ values of $\omega_{\textnormal{e}}$, in green $\omega_{\textnormal{e}}=0.8\omega_{\textnormal{m}}$ and orange $\omega_{\textnormal{e}}=1.2\omega_{\textnormal{m}}$. Normal mode splitting indicative of strong coupling might be observed with mechanical oscillators with quality factor as low as $Q_{\textnormal{m}}=\omega_{\textnormal{m}}/\Gamma_{\textnormal{m}}=40$ for a system with the characteristics described in the next section.

The solution to the set of coupled equations ${\bf Y}$ also suggest that the mechanical response is modified in the presence of an external driving voltage on the circuit $\tilde{V}_{\textnormal{ext}} (\omega)$. In a particular case, if the $LC$ resonator is thermally driven $\tilde{V}_{\text{ext}}(\omega)=\tilde{V}_{\textnormal{th}} (\omega)$ , this thermal drive can be measured with the mechanics. The response of the mechanical resonator to an external thermal voltage depends on a mechanical-voltage susceptibility $\chi_{\tiny \text{V}}(\omega)= \chi_{\text{eff}}(\omega)\sqrt{\frac{m}{L}}\frac{g_0 (\Gamma_{\textnormal{m}}-i \omega )}{ \left(i \Gamma_{\textnormal{e}} \omega +\omega^2 \omega_{\textnormal{e}}^2\right)}$. The PSD of the mechanics due to excitation in the electronics is then \begin{equation}
S_{{\tiny \text{V}} {\tiny \text{V}}}(\omega)= \langle |\chi_{\tiny \text{V}}(\omega)|^2\tilde{V}_{\textnormal{ext}} (\omega)\rangle.
\end{equation} 

The mechanical response to an external force and an electric drive is 
\begin{equation}\label{eq:z_totalspectrum}
\tilde{z}(\omega)
=\chi_{\text{eff}}(\omega)\tilde{F}_{\textnormal{ext}} (\omega) +  \chi_{\tiny \text{V}}(\omega)\tilde{V}_{\textnormal{ext}}(\omega).
\end{equation}
The mechanical spectrum of Eq. \eqref{eq:z_totalspectrum} will provide a way to experimentally measure the response of the mechanical oscillator to external forces applied on itself and external voltages applied on the $LC$, as well as its coupling.

\section{Magnetomechanical Device}\label{ExperimentalConsiderations}

Up to this point, we have treated the magnetomechanical system in a general fashion. Here we describe a design feasible to fabricate with  currently available photo and e-beam lithography techniques and materials. Here we describe some technical details regarding its fabrication and practical implementation. Our model considers the state-of-the-art experimental micro and nano fabrication techniques. In this particular design we study mostly the influence on the coupling rate $g_0$ due to the geometry and factors such as the height of the magnet $h_{\textnormal{m}}$, relative equilibrium vertical distance between the magnet and the coil $u_0$, number of turns of the inductor/coil $N$ and width $w$ of the wire/separation. The magnetomechanical system is in principle able to achieve coupling rates $g_0$ that exceed the values of the mechanical resonance frequency $\omega_{\textnormal{m}}$ which is extremely challenging for optomechanical systems.

\begin{figure}[h]
\includegraphics[scale=0.12]{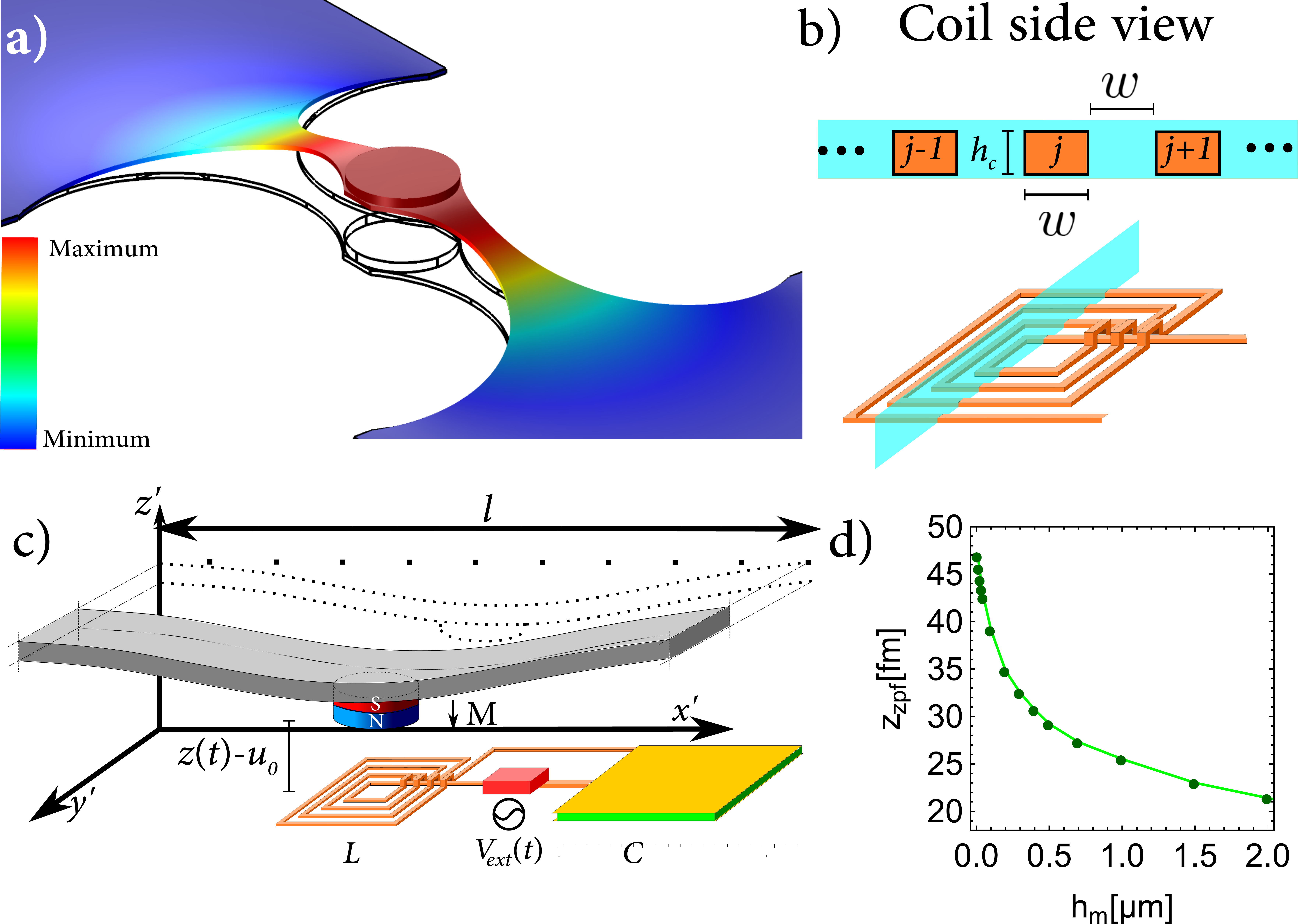}
\caption{\small a) Diagram of the fundamental mechanical mode of a double clamped beam made out of a Si$_3$N$_4$ membrane smoothly etched. The membrane has a cylinder on top that represents the magnet. b) Diagram of the coil on the electronic chip. In blue an axial plane cut showing the transversal area $w\times h_c$ of each wire of the coil. The width and spacing is $w$ and $h_c$ is the thickness of the wire. c) Representation of the flipped chip approach that places the double clamped beam resonator above the electronic chip. d) Plot of the zero point fluctuation $z_{\textnormal{\tiny ZPF}}$ for the fundamental mode as a function of the thickness/height of the magnet $h_{\textnormal{m}}$.}\label{fig:LCSchemeMembrane}
\end{figure}

The successful implementation of our magnetomechanical system (Fig. \ref{fig:LCSchemeMembrane}c) requires a two-chip fabrication process, separated in two main steps. The first chip (Fig. \ref{fig:LCSchemeMembrane}a) consist of a double clamped mechanical resonator (section \ref{Mechanical_System_fab}). The second chip (Fig. \ref{fig:LCSchemeMembrane}b) consist of a spiral coil and a planar capacitor fabricated on a sapphire substrate (section \ref{LC_Circuit_fab}). Each one of the chips is individually fabricated and later joint flipping the top chip (mechanical resonator) and adjusting the separation between them. These flipped joint chips form a system similar to the state-of-the-art 3D cavities recently developed \cite{yuan2015large,noguchi2016ground}.

\subsection{Mechanical system}\label{Mechanical_System_fab}

In this section we describe the protocol for the microfabrication of the mechanical oscillator of the magnetomechanical system. The mechanical oscillator could be fabricated as a double clamped beam 10$\mu$m long and $1\mu$m wide on a thin film Si$_3$N$_4$ membrane 100 nm thick on Si substrate. The membrane can be patterned using standard photolithography techniques, after exposure and development of the positive photoresist the open regions are etched using \textit{Reactive-Ion Etch} (RIE) fabrication technology. The finite element model (Comsol) in Fig. \ref{fig:LCSchemeMembrane}a shows  the fundamental motional mode shape of the loaded double clamped beam made out of Si$_3$N$_4$. It has been reported, Si$_3$N$_4$ has exceptional mechanical properties under cryogenic conditions \cite{purdy2012cavity}, which makes it suitable for future magnetomechanical setups. Mechanical oscillators made out of Si$_3$N$_4$ membranes have typical values for mechanical quality factor $Q_{\textnormal{m}}=\omega_{\textnormal{m}}/\Gamma_{\textnormal{m}}=10^{5}$.

On top of the patterned beam a second photolithographic step requires spin coating of a negative photoresist and expose it with the magnet pattern. After the exposure and development of the pattern a thin film magnetic material is deposited, similar to the coating process for cantilever's AFM magnetized tips. The final step is the gently release of the mechanical oscillator, which can be done using dry etch in a XeF$_2$ chamber for Si etch. This is an isotropic etch for Si which will remove the Si under the resonator. The thickness of the magnet $h_{\textnormal{m}}$ can be easily controlled during the deposition of the magnetic film. The cylindrical magnet is then formed at the center of the double clamped beam through lift-off of the negative resist. It is important to highlight that the remarkable mechanical properties of Si$_3$N$_4$ membranes remain largely unchanged when thin films are deposited on it \cite{yu2012control} far from the clamping region. The magnetic flux from the magnet will determine the magnitude of the interaction as the coupling rate $g_0\propto M$. It is desirable to have magnetic materials that support high density magnetization in thin films. A magnetic material with such characteristics and which has been extensively studied is Co-Fe, with a large number of different alloys \cite{dorsey1996cofe2o4}. Here we chose a standard one with density $\rho=7.81\text{g}/\text{cm}^3$ and a conservative value for the magnetization $\mu_0|{\bf M}|=0.264$T. Modern alloys have reached saturation magnetization up to $\mu_0|{\bf M}|=2.4$T \cite{fu_magnetically_2005}. The radius of the magnet is fixed to $r_{\textnormal{m}}$=0.5 $\mu$m and is a suitable size for photo or e-beam lithography, the only degree of freedom that we explore now is the height of the magnet $h_{\textnormal{m}}$ which is represented as the thickness of the magnetic thin film thickness. Considering the mass of the double clamped beam and the load of the magnet with its density, we calculate the effective mass $m$ and the resonance frequency $\omega_{\textnormal{m}}$ for the fundamental mechanical mode of the mechanical system as a function of $h_{\textnormal{m}}$, it is shown in Fig. \ref{fig:PlotsExperimental}a. By controlling the thickness of the deposited magnetic material, we can easily alter both $\omega_{\textnormal{m}}$, $m$ also on the coupling rate $g_0$. In Fig. \ref{fig:LCSchemeMembrane}d we plot the zero point motion $z_{\text{\tiny ZPF}}=\sqrt{\frac{\hbar}{2 m \omega_{\textnormal{m}}}}$ as a function of the magnet thickness $h_{\textnormal{m}}$. The control and tuning of the mechanical frequency has two different limits. In one limit, films which are only a few nanometers thick will result in a higher frequency mechanical oscillator, making the interaction with the $LC$ circuits technically more feasible. On the other hand, thicker films results in higher magnetic volumes and therefore stronger magnetic interactions. 


\subsection{Electrical circuit}\label{LC_Circuit_fab}
The electronic component of the magnetomechanical system requires to be fabricated on an individual chip.  Following standard nanofabrication techniques for coils \cite{teufel2011sideband, 
wollman2015quantum,
dieterle2016superconducting}, the chip can be fabricated on a sapphire substrate placing the micro/nano fabricated coil depicted in Fig. \ref{fig:LCSchemeMembrane}b. Where a first layer of metal is deposited on the surface of the chip. Following a spin coating of e-beam resist (PMMA) for later exposure and patterning of the central electrode and the flat part of the spiral inductor, etching the metal through wet etch. A sacrificial layer of resist is deposited and the exposed to pattern the bridge, following an oxide removal of a few nm with Ar bombarding on the surface and the successive  second layer of metal deposition to build the metallic bridge. A last step of resist removal either wet or dry needs to be implemented to remove the sacrificial layer.

A schematic representation of a spiral coil is shown in Fig. \ref{fig:LCSchemeMembrane}b, we also show  a transverse cut to define $w$ as the packing parameter. The packing parameter represents the width of the wire but also the spacing between each one of the wires that make a single turn. The maximum resolution of a nanofabrication system will determine the minimum value for $w$.

\subsection{Coupled system}

The coupled magnetomechanical system requires a double chip packaging. This particular packaging resembles the on chip 3D-cavity implementations, taking the chip with the mechanical oscillator and flipping it over the second chip with the $LC$ resonator as the scheme shows in the Fig. \ref{fig:LCSchemeMembrane}c.

\begin{figure}[h]
\includegraphics[scale=0.5]{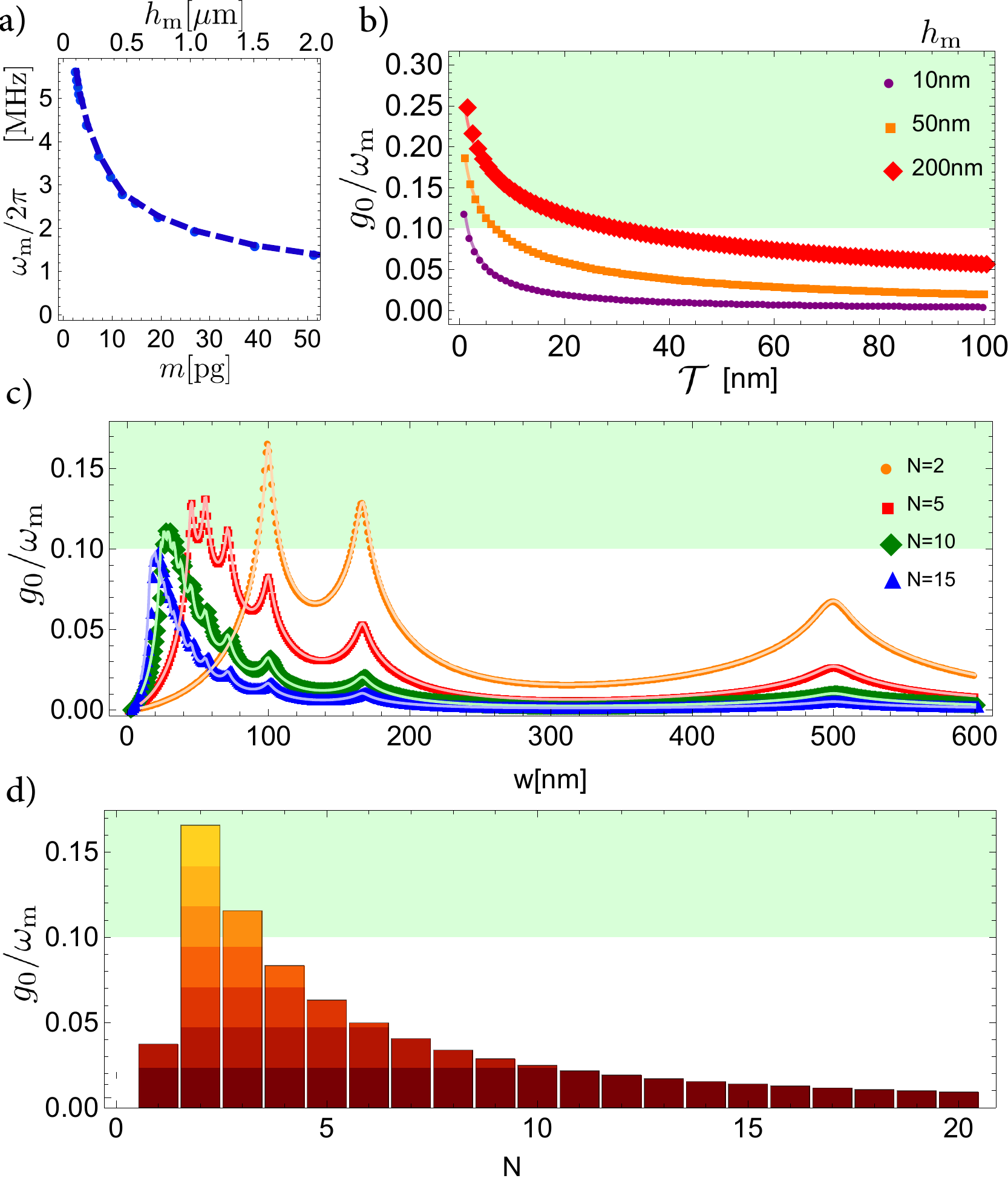}
\caption{\small a) Plot of $\omega_{\textnormal{m}}$ as a function of $h_{\textnormal{m}}$ and $m$ of the fundamental mode. b) Determination of the ratio $g_0/ \omega_{\textnormal{m}}$ as a function of $\mathcal{T}$ for $h_{\textnormal{m}}=10$nm, $h_{\textnormal{m}}=50$nm and $h_{\textnormal{m}}=200$nm. The parameters considered here are $w=$100 and $N=2$ of the inductor. In red the thickest magnet of $h_{\textnormal{m}}=$200 nm. In orange the curve that corresponds to $h_{\textnormal{m}}=$50 nm. In purple, the curve that corresponds to $h_{\textnormal{m}}=$10 nm. c) Estimation of $g_0/\omega_{\textnormal{m}}$ as a function of $w$. In these results $h_{\textnormal{m}}=200$ nm, with the one, the oscillator has an effective mass $m=9.9\times 10^{-15}$ kg and a mechanical resonant frequency $\omega_{\textnormal{m}}/2\pi=3.2$ MHz. The gap considered in this result is $\mathcal{T}=10$ nm. In blue, the curve corresponding to $N$=15, in green $N=$10, in red $N=$5 and orange $N=$2. d) Ratio $g_0/\omega_{\textnormal{m}}$ as a function of $N$. The numerical values were calculated considering a cylindrical magnet with $h_{\textnormal{m}}=$200 nm, $\mathcal{T}=10$ nm and $w=100$ nm. All the calculations were performed considering a magnetization $\mu_0 |M|=$0.264T. The blue shaded regions represent the values that lay in the ultra-strong coupling regime.}\label{fig:PlotsExperimental}
\end{figure}

Considering Eq.\eqref{eq:coupling} and the initial separation $u_0$, we can numerically estimate the value $g_0$ as a function of several parameters, such as the gap between the edge of the magnet $\mathcal{T}$, the number of turns of the coil $N$ and the packing parameter $w$ that represents the width and separation between the wires of the coil. The numerical results for the ratio of the coupling rate and the mechanical frequency $g_0/\omega_{\textnormal{m}}$ as a function of three different parameters are shown in Fig. \ref{fig:PlotsExperimental}a, \ref{fig:PlotsExperimental}b and \ref{fig:PlotsExperimental}c. In Fig. \ref{fig:PlotsExperimental}b the ratio $g_0/\omega_{\textnormal{m}}$ is presented as a function of the gap  $\mathcal{T}=u_0-h_{\textnormal{m}}/2$ between the coil and the edge of the magnet, for different thickness of the magnet $h_{\textnormal{m}}$ and with ($N$;$w$)=(2;100 nm). It is observable that the same tendency is followed for different thickness of the magnetic film. As $h_{\textnormal{m}}$ and the magnetic volume increases, the maximum of the coupling rate achievable increases, but not linearly. Fig. \ref{fig:PlotsExperimental}b shows a region in blue, where ratios of $g_0/\omega_{\textnormal{m}} \approx 0.1$, leads to physics in the ultra-strong coupling regime. The importance of this result relies on the unexplored regime for mechanical systems. This regime has been recently observed in superconducting qubits \cite{yoshihara_superconducting_2017}. Fig. \ref{fig:PlotsExperimental}c shows numerical simulations of $g_0/\omega_{\textnormal{m}}$ as a function of $w$ with parameters ($N$;$h_{\textnormal{m}}$;$\omega_{\textnormal{m}}$)=(2;200 nm; $2\pi\times3.2$ MHz). For this calculation we considered a magnet with $h_{\textnormal{m}}=200$ nm that for the fundamental mode has associated an effective mass $m=9.9\times 10^{-15}$ kg and whose mechanical resonant frequency is $\omega_{\textnormal{m}}/2\pi=3.2$MHz. This magnet is separated $\mathcal{T}=10$ nm from the spiral coil. We calculate the ratio $g_0/\omega_{\textnormal{m}}$ as a function of $w$ for different values of $N$. We observe that $N$ is also an important parameter due to its contribution to the inductance $L$. The inductance $L$ of the spiral square inductor was calculated with finite element methods software (Comsol) and compared with analytical expressions \cite{mohan1999simple}. The two methods yielded similar results and thus we chose to use the analytical expressions for simplicity and accuracy. The last parameter discussed in this paper is the enhancement of the coupling rate $g_0$ due to the number of turns of the nano fabricated coil. In Fig. \ref{fig:PlotsExperimental}d we plot the coupling rate ratio $g_0/\omega_{\textnormal{m}}$ as a function of $N$ keeping the parameters ($\mathcal{T}$;$h_{\textnormal{m}}$;$\omega_{\textnormal{m}}$)=(10 nm;200 nm; $2\pi\times3.2$ MHz) fixed while maximising over $w$. We observe that at $N=2$ the maximum ratio is obtained due to the low inductance which favours the increase   in the coupling. As it was described, the coupling rate $g_0\propto |{\bf M}|$, we have restricted our calculations to conservative magnetization values and consider that regimes such as deep-strong coupling can be achieved using modern alloys with larger magnetization.

\section*{Conclusion}

The magnetomechanical system that we have proposed in this work provides a suitable novel instrument to explore magnetomechanical dynamics in the strong and ultra-strong coupling regimes. 

We have introduced and described some physical effects such as magnetomechanical damping, or magnetomechanical frequency shift that might be further explored and implemented on cooling protocols, state swap, and electronic readout of the mechanical system. The magnetomechanical system provides an interface for novel hybrid quantum protocols on the control of mechanical oscillators using electric circuits. 

The interaction $\sim\hat{p}\hat{q}$ represents an attractive option for the implementation of novel protocols to perform back action evading measurements on mechanical oscillators via electronics. We estimated the number of excitations that populate the ground state and observe that the ground state of the magnetomechanical system is intrinsically entangled in the regime of low phonon-photon occupation regime. We also note that the Hamiltonian breaks time reversal symmetry due to its linear dependence on the mechanical momentum. We consider that this particular feature could help to understand some of the physics of symmetries at the mesoscale.

Considering the recent rapid progress in experimental techniques and fabrication processes such as photo and e-beam lithography, we consider that our magnetomechanical system is a feasibly proposal to be fabricated. We predict that a very large coupling $g_0$ might be potentially achieved in this fashion. This large coupling facilitates the implementation of already existing optomechanical protocols such as manipulation, control or cooling.

\section*{Appendix A: Magnetomechanical break of time reversal symmetry }\label{time_reversal}

The magnetomechanical system that we have introduced in this paper provides a diverse variety of interesting directions to explore quantum features for mechanical systems at the mesoscale. The linear character of the momentum coupling also represents an interesting framework to study the breaking of time reversal symmetry in this hybrid electromechanical interface.

In quantum mechanics, time-reversal symmetry is a bijective mapping of the Hilbert space. This mapping is symmetric if an only if it leaves all the observable probabilities invariant.  As we show below, the magnetomechanical linear Hamiltonian \eqref{eq:HamiltonianLinear}, is not invariant under time-reversal. The time reversal symmetry breaking phenomenon is a rare effect, which has been observed in circuit-QED \cite{koch2010time} but to our best knowledge it has not been observed in mechanical systems.

Following the definition for time-reversal symmetry we analyse $\hat{H}_L$, which is symmetric if and only if for a time-reversal operator $\hat{\Theta}$, there exist a phase $\vartheta(\hat{z})$, such that $\hat{H}_L=\hat{\Theta} \hat{H}_L \hat{\Theta}^{-1}$ is satisfied \cite{koch2010time}. The most relevant properties described by Koch \textit{et. al.} \cite{koch2010time} show that the operator $\hat{\Theta}$ acting on an eigenstate of the position $|\hat{z}\rangle $ leaves it invariant, but adding a phase $\Theta |\hat{z}\rangle = e^{i \vartheta(\hat{z})}|\hat{z}\rangle$ and the eigenstates of the position are time reversal symmetric $\hat{\Theta} \hat{z} \hat{\Theta}^{-1}=\hat{z} $. Under the same time reversal transformation $\hat{\Theta}$, the momentum is reflected and the gradient of a phase is added $\hat{\Theta} \hat{p}\hat{\Theta}^{-1} = -\hat{p} + \nabla \vartheta (\hat{z})$. The selection of the phase $\vartheta(\hat{z})$ is determined by the gauge choice $\varphi(z)$ discussed in Eq. \eqref{eq:Momenta_Gauge}. We apply the time reversal operator and obtain the transformed Hamiltonian, which reads as 
\begin{equation}
\begin{array}{cr}
\hat{\Theta} \hat{H}_L \hat{\Theta}^{-1}&=\displaystyle\frac{1}{2 m}\left(-\hat{p} +\sqrt{ L m}\ g_0 \hat{q} +\nabla \vartheta(\hat{z})\right)^2
\\
\\
&
+
\displaystyle\omega_{\textnormal{m}}^2\frac{ m \hat{z}^2}{2}
+
\displaystyle \frac{\hat{\phi}^2}{2L}
+
\omega_{\textnormal{e}}^2\frac{ L  \hat{q}^2}{2}.
\end{array}
\end{equation}
The condition $\hat{H}_L=\hat{\Theta} \hat{H}_L \hat{\Theta}^{-1}$ to identify this system as time reversal symmetric implies that a solution is given by $\nabla \vartheta(\hat{z})=-2 g_0\ \sqrt{L m} \hat{q}$. In the particular case when the coupling is absent $g_0=0$ the solution to this condition is satisfied and the Hamiltonian satisfies the time reversal symmetry $\hat{H}_L=\hat{\Theta} \hat{H}_L \hat{\Theta}^{-1}$. In any other case $g_0\neq 0 $ the Eq.\eqref{eq:Momenta_Gauge} suggest that the gauge choice is related to an electromagnetic auxiliary field ${\bf A}=\nabla \varphi(\hat{z})$ therefore it must satisfy $\nabla \times{\bf A}= {\bf 0}$, which in the presence of a magnetic field clearly contradicts ${\bf B}=\nabla \times{\bf A}$.

Summarising, the magnetomechanical system described in this paper satisfies  time reversal symmetry in the absence of coupling, but breaks it when the linear magnetomechanical coupling is present. We want to highlight this property for the magnetomechanical system as a an alternative to explore time reversal symmetry breaking for mechanical systems.



\bibliography{PRA_Library}
\end{document}